hep-ph/9308366
\documentstyle[12pt]{article}

\newlength{\dinwidth}
\newlength{\dinmargin}
\newcommand{\resection}[1]{\setcounter{equation}{0}\section{#1}}

\begin{document}
\vspace*{3cm}
\begin{center}
  \begin{Large}
  \begin{bf}
ONE LOOP CALCULATION OF THE \\
$\epsilon_3$ PARAMETER\\
 WITHIN THE EXTENDED BESS MODEL\\
  \end{bf}
  \end{Large}
  \vspace{2cm}
  \begin{large}
R. Casalbuoni, S. De Curtis and M. Grazzini\\
  \end{large}
\vspace{4mm}
Dipartimento di Fisica, Universit\`a di Firenze, Largo E. Fermi, 2 - 50125
Firenze (Italy)\\
I.N.F.N., Sezione di Firenze, Largo E. Fermi, 2 - 50125
Firenze (Italy)\\
\end{center}
  \vspace{4cm}
\begin{center}
  \begin{bf}
  ABSTRACT
  \end{bf}
\end{center}
  \vspace{5mm}
\noindent
The existence of a strongly interacting sector responsible for the electroweak
symmetry breaking is assumed. As a consequence vector and axial-vector
bound states may be formed. These resonances mix with
the Standard Model gauge bosons and are of primary phenomenological
importance for the LEP physics.
The extended BESS model is an effective scheme based on the symmetry group
$SU(8)_L\otimes SU(8)_R$,
describing in a consistent way the interactions among the pseudo-Goldstone
bosons, vector and axial-vector resonances and the standard gauge bosons.
In a previous paper, the contribution from extended BESS to the electroweak
oblique corrections was evaluated. However, only an estimate of the effects
coming from mass and wave function renormalization of the new resonances,
was given. Here we complete the evaluation by computing explicitly these
effects. We confirm the previous result, that is, in spite of the great
precision of the present LEP measurements, the extended
BESS parameter space is not
very much constrained.

\newpage
\newcommand{\f}[2]{\frac{#1}{#2}}
\def\lq{\left [}
\def\rq{\right ]}
\def\dmu{\partial_{\mu}}
\def\dmus{\partial^{\mu}}
\def\gp{g'}
\def\gs{g''}
\def\ggs{\frac{g}{\gs}}
\def\eps{{\epsilon}}
\newcommand{\be}{\begin{equation}}
\newcommand{\ee}{\end{equation}}
\newcommand{\bea}{\begin{eqnarray}}
\newcommand{\eea}{\end{eqnarray}}
\newcommand{\nn}{\nonumber}
\newcommand{\dd}{\displaystyle}
\setcounter{page}{1}
\resection{Introduction}

Precision measurements are now giving serious restrictions on
the possibility of a new strong interacting sector being at the origin of
the symmetry breaking in the electroweak theory.
A natural consequence of such a strong sector is the occurrence of resonances
in the $TeV$ region. Among them, spin-1 resonances would be of particular
interest as they would, already through mixing effects, affect the
self-energies
of the standard model gauge bosons.

An effective  scheme (BESS model ref. \cite{bess}),
 introduced  several years ago, describes a minimal version of such a scenario
by introducing a triplet of vector resonances and no Goldstone bosons, being
based on a non linear $SU(2)_L \otimes SU(2)_R$ chiral model.

If the strongly interacting sector gives rise to pseudo-Goldstone bosons, it
has been shown \cite{pseudo} that their loop contribution to the vector
boson self-energies is not negligible.
For this reason, in ref. \cite{bessu8}, an extended version of the BESS model,
based on a chiral $SU(8)_L\otimes SU(8)_R$, was given.

The model contains 60 pseudo-Goldstone bosons in the spectrum and, as a
further generalization, it describes also axial-vector resonances together
with vector ones.
The physics of this model,  as far as the LEP experiments are concerned,
is such that it gives corrections only to the self-energies of the standard
electroweak gauge bosons.
These corrections have been evaluated in ref. \cite{bessu8} at one-loop level
by defining the theory with a cut-off $\Lambda$.
In that calculation the contributions coming from mass and wave function
renormalization of the new resonances were evaluated in an approximate
way using a dispersive representation.
Here we want to give a more complete analysis by  including in the calculation
all the one-loop diagrams contributing to self-energies of
the new vector and axial-vector
resonances.

\resection{Gauge boson self-energies within the extended BESS model}

The extended BESS model effective Lagrangian, as introduced in ref. [3],
describes
vector and axial-vector gauge bosons interacting
with the Goldstone bosons and the
Standard Model (SM) vector bosons.

The effective Lagrangian is obtained by enlarging
the symmetry group of the BESS model based on a non-linear $SU(2)_L\otimes
SU(2)_R$ $\sigma$ model as described in ref. \cite{bess}, to
contain a "hidden" $SU(8)_L\otimes SU(8)_R$
gauge group, by introducing the covariant derivatives
with respect to the gauge fields associated to the "hidden" symmetry and with
respect to the SM gauge fields, and by  adding  together
all the independent invariant terms containing at
most two derivatives.

The new resonances $V^A$ and $A^A$  ($A$=1,...,63)
are described as Yang-Mills fields which acquire
mass through the same symmetry breaking mechanism which gives mass to the
ordinary gauge bosons.

Working  in the unitary gauge for the $V$,  $A$ and
SM gauge bosons leaves us with
60 Goldstones as physical particles in the spectrum.
We will assume that they acquire mass through radiative corrections.

An important property is that
the mixing terms among the SM gauge bosons and
the new vector and axial-vector resonances
are the same as in the axial-vector
extension of the $SU(2)$-BESS model \cite{assiali} plus an extra term
in the neutral sector involving the singlet (under $SU(3)_c\otimes
SU(2)_L\otimes U(1)_Y$) field $V_D$.
The neutral Lagrangian which gives the mixing among the new resonances and the
standard $SU(2)_L\otimes U(1)_Y$ gauge bosons is
\bea
{\cal L}^{mix} &=&
\f{v^2}{8}\Big[ a(g W^3-\gp Y)^2+b(g W^3+\gp Y-\gs V^3)^2\nn\\
& & +b (\f{2}{\sqrt{3}}
  \gp Y-\gs V_D)^2+c(g W^3-\gp Y+\gs A^3)^2+d\gs^2 (A^3)^2\Big]
\eea
where
$a,~b,~c,~d$ are free parameters, $g$ and $\gp$
are the gauge  coupling constants
of $W$ and $Y$ respectively, and $\gs$ is the self
coupling of the new gauge bosons.
{}From eq. (2.1) we see that the mixing
terms are proportional to $x\equiv g/\gs$ so one can
decouple the new resonances from the SM gauge bosons,
by taking
the large $\gs$ limit (small $x$). The SM results are recovered by
putting $x=0$ and
fixing  the normalization $a+cd/(c+d)=1$.

By diagonalizing the mass matrix it turns out that
the SM gauge boson masses get corrected by terms of the order of
$(g/\gs)^2$ (the photon remains massless)
while the vector bosons (and the axial ones)
are degenerate if one neglects the
weak corrections, with their masses
proportional to $\gs^2$.

In ref. \cite{bessu8} we have explicitly evaluated the interaction terms
among the SM gauge bosons, the new resonances and the pseudo-Goldstone bosons
(PGB). In the following application, we will need also the kinetic terms
for the $V$ and $A$ resonances which are triplets under $SU(2)$. The explicit
 expression is given in ref. \cite{self}.

The BESS model parameter space is four-dimensional;
we will choose, as free parameters,
the masses of the vector and the axial-vector resonances $M_V$ and $M_A$,
their gauge coupling constant $\gs$ and
$z=c/(c+d)$, which measures the ratio of the mixings
$W-A$ and $W-V$.

The mixing described in (2.1)
induces corrections to the self-energies of the SM.
We define the scalar part of the SM vector boson self-energies through the
relation
\be
\Pi_{ij}^{\mu\nu}(p^2)=-i\Pi_{ij}(p^2)g^{\mu\nu}+p^\mu p^\nu~terms
\ee
where the indices $i$ and $j$ run over the ordinary gauge vector bosons.

In the neutral sector we choose to work with the SM fields ($\theta$ is the
Weinberg angle)
\bea
& &W^3 =c_\theta~ Z+s_\theta~ \gamma\\
& & Y =-s_\theta~ Z+c_\theta~ \gamma
\eea

The corrections from the vector and axial-vector resonances and from the
PGB of extended BESS are purely oblique and only affect
the scalar gauge boson
self-energy terms $\Pi_{WW}$, $\Pi_{33}$, $\Pi_{30}$, and $\Pi_{00}$.

It is then convenient to introduce the following combinations
(see ref. \cite{altarelli}):
\bea
\epsilon_1&=&\frac{\Pi_{33}(0)-\Pi_{WW}(0)}{M^2_W}\\
\epsilon_2&=&\frac{\Pi_{WW}(M^2_W)-\Pi_{WW}(0)}{M^2_W}-
\frac{\Pi_{33}(M^2_Z)-\Pi_{33}(0)}{M^2_Z}\\
\epsilon_3 &=&\frac {c_\theta}{s_\theta}
\frac{\Pi_{30}(M^2_Z)-\Pi_{30}(0)}{M^2_Z}
\eea

In ref. \cite{bessu8} we have shown that the tree-level
contribution to the $\eps_i$ parameters coming from the mixing
of the $V$ and $A$ bosons with the $W$, $Z$ and $\gamma$ is the same we found
for $SU(2)$-BESS model in ref. \cite{self}, or
\be
\epsilon_1=0~~~~\epsilon_2\simeq 0,~~~~\epsilon_3\simeq x^2 (1-z^2)
\ee
where the last two results were obtained for large $M_{V,A}$
($M_{V,A}>>M_{W,Z}$).
The reason is that
the new $V_D$ boson mixes only with $Y$ (see (2.1)) and, as a consequence,
it does not affect $\Pi_{30}$ and it contributes only to the definition
of the electric charge and $M_Z$.
This means that our result can be extended to
a $SU(N)_L\otimes SU(N)_R$ model, in fact
in these models the new fields associated to the
diagonal generators will be mixed only with the hypercharge field $Y$.

In addition to the self-energy corrections arising from the mixing, one has
loop
corrections.
In ref. \cite{bessu8} we have implemented the calculation of the $\eps_i$
parameters at one-loop level.
Being BESS a non renormalizable model,
the loop integrals were evaluated using a cut-off $\Lambda$.
In Fig. 1 we list the graphs contributing to $\eps_3$ at one-loop
level.
In ref. \cite{bessu8} the calculation was performed
neglecting the contributions to $\eps_3$ coming from the self-energies
of the new resonances (which are schematically represented in Fig. 1 by the
first two graphs) and
we presented a simple estimate of these contributions
through  dispersive integrals.
In particular
the first two graphs in Fig. 1 represent the propagators for the $V^3$
and $A^3$ gauge bosons, corrected and opportunely
renormalized at one-loop level, which must be inserted, via mixing,
in the $\Pi_{30}$ function.
Here we will give the complete one-loop result for $\eps_3$.

\resection{One-loop evaluation of $\epsilon_3$}

The graphs contributing to the $V^3$ and $A^3$ self-energies are listed
in Fig. 2. They have been evaluated using the technique of
dimensional regularization. We have not included the tadpoles because they give
a constant (as function of $p^2$) contribution to the self-energies
 and so they can be disposed of by
mass renormalization. Furthermore, being $\eps_3$ an isospin symmetric
observable,  we have assumed a common
non-vanishing mass $M_\pi$ for all the PGB.

The first contribution to the $A_3$ self-energy, given by the graph in
Fig. 2$a$, is

\bea
\Pi_A^{(a)}(p^2)&=&
15~\f{g_{VA\pi}^2}{16\pi^2}\Big[\Big(\f{3}{4}-\f{1}{4}
\f{M^2_\pi}{M^2_V}+\f{p^2}{12M^2_V}\Big)\Big(\ln\f{\Lambda^2}{M_V M_\pi}
-\gamma\Big)\nn\\
& &+\f{M^2_V-M^2_\pi}{2p^2}\Big(1+\f{(M_\pi^2-M^2_V)^2}{12p^2M^2_V}
-\f{M^2_V+M_\pi^2}{4M^2_V}\Big)\ln\f{M^2_\pi}{M^2_V}\nn\\
& &+\Big(\f{1}{3}\f{M^2_\pi}{M^2_V}-\f{5}{3}-\f{1}{6}\f{p^2}{M^2_V}
-\f{(M^2_\pi-M^2_V)^2}{6M^2_V p^2}\Big)~\beta(M^2_V,M^2_\pi,p^2)
{}~\ln l(M_V,M_\pi,p^2)\nn\\
& &+\f{17}{12}-\f{7}{12}\f{M^2_\pi}{M^2_V}+\f{2}{9}\f{p^2}{M^2_V}
+\f{(M^2_\pi-M^2_V)^2}{12M^2_V p^2}\Big]
\eea
where the factor 15 gives the number of the pairs $V\pi$ circulating in the
loop (we are working in the unitary gauge where the $SU(2)$ triplet $\pi^a$
is eaten up by $W^\pm,Z$) and
\bea
\beta(M_1^2,M^2_2,p^2)&=&\f{\sqrt{(M^2_1-M^2_2)^2-2(M^2_1+M^2_2)p^2+p^4}}
{{2p^2}}\\
l(M_1,M_2,p^2)&=&
\f{s_+(M_1,M_2,p^2)-s_-(M_1,M_2,p^2)}{s_+(M_1,M_2,p^2)+s_-(M_1,M_2,p^2)}\\
s_\pm(M_1,M_2,p^2)&=&\sqrt{(M_1\pm M_2)^2-p^2}
\eea
$\gamma$ is the Euler's constant $(\simeq 0.577)$,
and $g_{VA\pi}$ is a trilinear coupling given in ref. \cite{bessu8}
\be
g_{VA\pi}= \f{v}{4}z\gs^2\f{x^2}{r_V}\left(\f{r_V}{r_A}-1\right)
\ee
with $v\simeq 246~GeV$ and $r_{V,A}\simeq M_W^2/M_{V,A}^2$
 in the large $\gs$ limit.

The second contribution to $\Pi_A$, given by the graph in Fig. 2$b$, is
\bea
\Pi_A^{(b)} (p^2) &=& 16~\f{\gs^2}{16 \pi^2} \Big[ a(M^2_A,M^2_V,p^2)
+b(M^2_A,M^2_V,p^2)\Big(\ln\f{\Lambda^2}{M_A M_V}-\gamma\Big)\nn\\
& &+c(M^2_A,M^2_V,p^2)
{}~\beta(M^2_A,M^2_V,p^2)~\ln
l(M_A,M_V,p^2)\nn\\
& &+d(M_A^2,M_V^2,p^2) \ln \f{M_V^2}{M^2_A}\Big]
\eea
where, again the factor 16 is a multiplicity factor, and
\bea
a(x_1,x_2,t)&=&\f{1}{72x_1 x_2 t}\Big(-405 x_1 x_2 (x_1+x_2)t+90(x_1^3+x_2^3)t
+6f(x_1,x_2,t)\nn\\
& &-452x_1 x_2 t^2-182(x_1^2+x_2^2)t^2+16 t^4+70(x_1+x_2)t^3\Big)\\
b(x_1,x_2,t)&=&\f{1}{12x_1 x_2} \Big(9(x_1^3+x_2^3)-36x_1 x_2(x_1+x_2)
-17  (x_1^2+x_2^2)t\nn\\
& &-50 x_1 x_2 t+7(x_1+x_2)t^2+t^3 \Big)\\
c(x_1,x_2,t) &=& \f{1}{6x_1x_2t}
\Big(-f(x_1,x_2,t)-8(x^3_1+x_2^3)t+32 x_1 x_2 (x_1+x_2)t
+18(x_1^2+x_2^2)t^2\nn\\
& &+32 x_1 x_2 t^2-8 (x_1+x_2)t^3-t^4\Big)\\
d(x_1,x_2,t)&=&\f{x_1-x_2}{24 x_1 x_2 t^2} \Big( f(x_1,x_2,t)+7(x_1^3+y^3_2)t
-43 x_1 x_2 (x_1+x_2)t-17(x_1^2+x_2^2)t^2\nn\\
& &-53x_1 x_2 t^2+9(x_1+x_2)t^3\Big)\\
f(x_1,x_2,t)&=&(x_1-x_2)^2(x_1^2+10 x_1 x_2+x_2^2)
\eea

The contribution to the $V^3$ self-energy coming  from the graph in
Fig. 2$c$ can be evaluated from $\Pi_A^{(a)}$ with the substitution
$M_V\to M_A$.

{}From the graph in Fig. 2$d$ we get
\bea
\Pi_V^{(d)}(p^2)&=&15~
\f{g^2_{V\pi\pi}}{8\pi^2}\Big[\Big(\f{p^2}{6}-M^2_\pi\Big)\Big(
\ln\f{\Lambda^2}{M^2_\pi}-\gamma\Big)\nn\\
& &+\f{8}{3}p^2 \alpha^3(M^2_\pi,p^2) \arctan\f{1}{2\alpha (M^2_\pi,p^2)}
-\f{7}{3}M^2_\pi+\f{4}{9}p^2\Big]
\eea
where, again, the factor 15 is a multiplicity factor
\be
\alpha(M^2_\pi,p^2)=\sqrt{\f{M_\pi^2}{p^2}-\f{1}{4}}
\ee
and the trilinear coupling $g_{V\pi\pi}$ (as given in ref. \cite{bessu8}) is
\be
g_{V\pi\pi}=\f{\gs}{4}\f{x^2}{r_V}(1-z^2)
\ee

Finally, the contributions corresponding to the graphs in Figs. 2$e$ and 2$f$
can be evaluated directly from $\Pi_A^{(b)}$ with the substitutions
$M_A\to M_V$ and $M_V\to M_A$ respectively.

We now introduce counterterms in order to get canonical propagator at the
mass of the resonance
\be
\Pi^{ren}_i(p^2)= \Pi_i(p^2)+p^2(Z_{3i}-1)+Z_i
\ee
with $i=A,~V$. The renormalization conditions are
\be
\left\{
\begin{array}{lcc}
Re \Pi^{ren}_i(p^2)\Big|_{p^2=M^2_i}=0~~~~~~~~& &\\
\nn\\
{\displaystyle \lim_{p^2\rightarrow M^2_i}  \f{p^2-M^2_i}
{p^2-M^2_i+Re\Pi^{ren}_i(p^2)}}=1 & &
\end{array}
\right.
\ee
where we have taken the real part of $\Pi^{ren}_i$ because the vacuum
polarization tensor
develops an imaginary part above threshold for the possible decay processes
of $A^3$ and $V^3$.
By substituting (3.15) in (3.16) we get
\be
Z_{3i}-1=-\f{d}{dp^2}Re \Pi_i(p^2)\Big|_{p^2=M^2_i}
\ee
\be
Z_i=-Re\Pi_i(M_i^2)-M_i^2(Z_{3i}-1)
\ee

We are now able to compute the first two contributions to $\Pi_{30}$
shown in Fig. 1. In fact, one has only to substitute the expressions for
$\Pi^{ren}_A(p^2)$ and $\Pi^{ren}_V(p^2)$ in the following
relation (see ref. \cite{self})
\be
\Pi_{30}^{self}(p^2)=
-x^2 \f{s_\theta}{c_\theta} p^2\Big(\f{M^2_V}{p^2-M^2_V+
\Pi^{ren}_V(p^2)}
-z^2 \f{ M^2_A}{p^2-M^2_A+
\Pi^{ren}_A(p^2)}\Big)
\ee
and then, to use it in eq. (2.7) for the calculation of $\eps_3^{self}$.
The value of $\eps_3^{self}$ predicted by the model is real
because $p^2=M^2_Z$ is below threshold. Notice that by considering
$\Pi^{ren}_A(p^2)=\Pi^{ren}_V(p^2)=0$  in (3.19), one recovers,
for large $M_{V,A}$, the tree level
value of  $\eps_3$ as given in eq. (2.8).

The contribution to $\eps_3$ of the remaining graphs of Fig. 1 was calculated
in ref. \cite{bessu8}:
\bea
\epsilon_3^{loop} &\simeq &
\frac{g^2}{16\pi^2} \frac{5}{8}\Big \{\left
(\log\frac{\Lambda^2}{M_\pi^2}-\gamma\right)
\Big[2-\frac{1}{2} \frac{x^4}{r_V^2}(1-z^2)^2\nn\\
& &-\frac{x^2}{r_V}(1-z^2\frac{r_V}{r_A})
 \Big(1 -z^2 -z^2(1-\frac{r_V}{r_A})\Big)-z^2
\frac{x^2}{r_A}(1-\frac{r_A}{r_V})^2\Big]\nn\\
& &-\frac{x^2}{r_V}(1-z^2\frac{r_V}{r_A})
\Big(1 -z^2 -z^2(1-\frac{r_V}{r_A})\Big)\Big(A(M_\pi^2,M_V^2)+1\Big)\nn\\
& &-z^2\frac{x^2}{r_A}(1-\frac{r_A}{r_V})^2\Big(A(M_\pi^2,M_A^2)+1\Big)
\Big\}
\eea
where
\bea
A(M_\pi^2,M^2)&=&
\frac{M^6+9  M^4 M_\pi^2}{(M^2-M_\pi^2)^3}
\log{\frac{M_\pi^2}{M^2}}\nn\\
& &+\frac{1}{6 (M^2-M_\pi^2)^3} (M_\pi^6-27 M_\pi^4 M^2-9 M_\pi^2 M^4
+35 M^6)
\eea

Furthermore, the numerical analysis has to include electroweak radiative
corrections ($\eps_3^{rad}$). Our assumption, physically plausible in the
presence of
new degrees of freedom related to a larger scale, is to take the usual one-loop
radiative corrections of the SM  with the
identification of the Higgs mass as the cut-off $\Lambda$ which regularizes
BESS
at high momenta. Finally we have
\be
\eps_3=\eps_3^{rad}+\eps_3^{self}+\eps_3^{loop}
\ee

\resection{Numerical results}

We now want to evaluate numerically the $\eps_3$ parameter as predicted by the
 extended BESS model.
In order to reduce the parameter space, we will assume, as we did in ref.
\cite{bessu8}, the validity of the Weinberg sum rules (WSR) \cite{losecco}.
In this way we can eliminate two parameters, for example $M_A$ and $\gs$ in
favour of $M_V$ and $z$. We get \cite{bessu8}
\be
M_A^2=\f{M_V^2}{z}~~~~~~~~\gs=2\f{M_V}{v}\sqrt{1-z}
\ee
with $0<z<1$.
Then, the free parameters of our analysis are: $M_V$, $z$, $M_\pi$ and
$\Lambda$.

Using the most recent results from LEP experiments combined with the low
energy weak data \cite{lep}
\be
\eps_3=(1.3\pm 3.1)\times 10^{-3}
\ee
we can derive the bounds on the parameter space of the extended
BESS model coming from the estimate of the corrections to $\eps_3$.

Concerning the electroweak radiative corrections, following ref. \cite{cara},
we use
\be
\eps_3^{rad}=0.00696
\ee
corresponding to $m_{top}=150~GeV$ and $M_H=\Lambda=1.5~TeV$.

In Fig. 3 we give the 90\% C.L. allowed region in the plane $(z,M_V)$ for
two values of $M_\pi/\Lambda$ and $\Lambda=1.5~TeV$.
The effects of including the contribution coming from the $V$ and $A$
self-energies are generally small and
go in the direction of enlarging the allowed region.
This result is in accordance with the observations done in ref. \cite{bessu8}
that the $V$ and $A$
self-energy contributions are generally negative and decrease
the value of $\eps_3^{tree}$ of the tree level.
The corrections due to the self-energies are more sizeable in
the region close to $z=1$.
In fact, by using the WSR, the expression
of $\eps_3^{loop}$ in eq. (3.20) vanishes in the limit $z\to 1$.
The reason for this is twofold. First, the contributions
from the three graphs with PGB loops (see Fig. 1) cancel among themselves,
second, the WSR imply that for $z=1$ also
$r_V=r_A$  corresponding to a complete degeneration among $V$ and $A$
resonances (notice also that $\eps_3^{tree}=0$ for
$z=1$). This latter property is not
 true anymore for the complete expression of $\eps_3$
since the $V$ and $A$ resonances
get different corrections from  the self-energies, also at $z=1$.

In Fig. 4 we give the 90\% C.L. allowed region in the plane $(M_V,M_\pi)$ for
two values of $z$ and $\Lambda=1.5~TeV$. We notice that for $M_\pi>100~GeV$
the bounds do not depend very much  on the mass of the PGB and that the
allowed region shrinks for increasing values of $z$.

Finally, for completeness, in Fig. 5 we also give the 90\% C.L.
allowed region in the plane $(z,M_\pi)$
for two values of $M_V$ and $\Lambda=1.5~TeV$.

{}From Figs. 3 and 4 we see that there is an absolute lower
bound on $M_V$, independent on $z$ and $M_\pi$,
of about 900 $GeV$.

In conclusion the calculation we have done here shows that the approximation
used in ref. \cite{bessu8} for the evaluation of the self-energies of the
new resonances was essentially correct. However it should be noticed that
a sizeable difference arises in the region close to $z=1$ where, as explained
before, the different one-loop contributions  for the vector and the
axial-vector resonances forbid the cancellation arising from the other
diagrams.

\newpage

 \begin{center}
  \begin{Large}
  \begin{bf}
  Figure Captions
  \end{bf}
  \end{Large}
  \end{center}
  \vspace{5mm}
\begin{description}
\item [Fig. 1] Graphs contributing to $\eps_3$ at one-loop level
               within the extended BESS model.
\item [Fig. 2]  $A^3$ and $V^3$ self-energy contributions.
\item [Fig. 3]  90\% C.L. allowed regions in the plane $(z,M_V)$ from
        	$\epsilon_3$ for $\Lambda=1.5~TeV$.
	        The solid (dashed) line is the lower bound coming from
		the total one-loop effect for $M_\pi/\Lambda=0.10$
                ($M_\pi/\Lambda=0.35$).
\item [Fig. 4]  90\% C.L. allowed regions in the plane $(M_V,M_\pi)$ from
        	$\epsilon_3$ for $\Lambda=1.5~TeV$.
	        The solid (dashed) line is the bound coming from
		the total one-loop effect for $z=0.2$
                ($z=0.8$). The allowed regions lie on the right
		of the curves.
\item [Fig. 5]  90\% C.L. allowed regions in the plane $(z,M_\pi)$ from
        	$\epsilon_3$ for $\Lambda=1.5~TeV$.
	        The solid (dashed) line is the bound coming from
		the total one-loop effect for $M_V=1~TeV$
                ($M_V=1.5~TeV$). The allowed regions lie on the left
		of the curves.

\end{description}

\begin{thebibliography}{99}

\bibitem{bess}
R.Casalbuoni, S.De Curtis, D.Dominici and R.Gatto,
       Phys. Lett. {\bf B155}  (1985) 95; $ibidem$
       Nucl. Phys. {\bf B282} (1987) 235

\bibitem{pseudo}
M.Golden and L.Randall, Nucl. Phys. {\bf B361} (1991) 3;
R.Cahn and M.Suzuki, Phys. Rev. {\bf D44} (1991) 3641;
M.E.Peskin and T.Takeuchi, Phys. Rev. {\bf D46} (1992) 381


\bibitem{bessu8}
   R.Casalbuoni, S.De Curtis, A.Deandrea, N.Di Bartolomeo, R.Gatto,
   D.Dominici and F.Feruglio, UGVA-DPT 1992/07-778, July 1992

\bibitem{assiali}
R.Casalbuoni, S.De Curtis, D.Dominici, F.Feruglio and R.Gatto,
    Int. Journ. of Mod. Phys. {\bf A4}  (1989) 1065

\bibitem{self}
   R.Casalbuoni, S.De Curtis, D.Dominici, F.Feruglio and R.Gatto,
   Phys. Lett. {\bf B258}  (1991) 161


\bibitem{altarelli}
   G.Altarelli and R.Barbieri, Phys. Lett. {\bf B253} (1991) 161;
   D.C.Kennedy and P.\break
   Langacker, Phys. Rev. Lett. {\bf 65} (1990) 2967; Erratum
   {\bf 66} (1991) 395; $ibidem$ Phys. Rev. {\bf D44} (1991) 1591;
   A.Ali and G.Degrassi, DESY-91-035, to be published in the {\it M.A.B. B\'eg
   Memorial Volume} (World Scientific, Singapore, 1991);
   G.Altarelli, R.Barbieri and S.Jadach, Nucl. Phys. {\bf B369} (1992) 3

\bibitem{losecco}
 S.Weinberg, Phys. Rev. Lett. {\bf 18} (1967) 507;
 C.Bernard, A.Duncan, J.Lo Secco and S.Weinberg, Phys. Rev. {\bf D12}
 (1975) 792

\bibitem{lep}
 G.Altarelli, talk given at the Conference "Incontro sulla Fisica a LEP"
 Florence, April 1-2, 1993

\bibitem{cara}
 G.Altarelli, R.Barbieri and F.Caravaglios, CERN-TH 6770/93, January 1993

\end{thebibliography}
\end{document}